\documentclass[twocolumn]{revtex4-1}
\usepackage{bm}
\usepackage[pdftex]{graphicx}
\usepackage{latexsym}
\usepackage{epsfig}
\usepackage{amsmath,amssymb}
\usepackage{amsfonts}
\usepackage{palatino}
\usepackage[latin1]{inputenc}
\usepackage{epsf}
\usepackage{latexsym}
\usepackage{calc}
\usepackage{color}
\usepackage{shadow}
\usepackage{epsfig}
\usepackage{float}
\usepackage[all]{xy}
\usepackage[pdftex, pdfstartview = {FitH}]{hyperref}
\usepackage{csquotes}
\usepackage{afterpage}

\usepackage{braket}

\begin{document}
\title{Exact Exchange: a pathway for a Density Functional Theory of the Integer Quantum Hall Effect}

\author{D. Miravet}
\affiliation{Centro At\'omico Bariloche, CNEA, CONICET, 8400 S. C. de Bariloche, R\'io Negro, Argentina}
\author{G. J. Ferreira}
\affiliation{Instituto de F\'isica, Universidade Federal de Uberl\^andia, Uberl\^andia, Minas Gerais 38400-902, Brazil}
\author{C. R. Proetto}
\affiliation{Centro At\'omico Bariloche and Instituto Balseiro, 8400 S. C. de Bariloche, R\'io Negro, Argentina}

%\pacs{71.15.Mb}{Density functional theory}
%\pacs{73.43.-f}{Quantum Hall effects}
%\pacs{73.21.Fg}{Quantum wells}

\begin{abstract}
It is shown here that the Exact Exchange (EE) formalism provides a natural and rigorous approach for a Density Functional Theory (DFT) of the Integer Quantum Hall Effect (IQHE). Application of a novel EE method to a quasi two-dimensional electron gas (q2DEG) subjected to a perpendicular magnetic field leads to the following main findings. \textit{i)} the microscopic exchange energy functional of the IQHE has been obtained, whose main feature being that it minimizes with a discontinuous derivative at every integer filling factor $\nu$; \textit{ii)} an analytical solution is found for the magnetic-field dependent EE potential, in the one-subband regime; \textit{iii)} as a consequence of \textit{i)}, the EE potential display sharp discontinuities at every integer $\nu$; and \textit{iv)} the widely used Local Spin Density Approximation (LSDA) is strongly violated for filling factors close to integer values.
\end{abstract}

\maketitle

\section{Introduction}
Density Functional Theory (DFT) is one of the most used computational tools for the theoretical study of inhomogeneous interacting systems such as atoms, molecules, and solids\cite{PY89}. In its Kohn-Sham (KS) implementation, the real interacting electronic system is rigorously mapped into an auxiliary non-interacting system, where all the interactions are included in an effective single-particle potential, the spin-dependent KS potential $v_{\text{KS}}^{\sigma}(\mathbf{r})$. The crucial ingredient of this potential is its exchange-correlation (xc) contribution, that is obtained from the xc energy functional $E_{\text{xc}}[\rho(\mathbf{r})]$, with $\rho(\mathbf{r}) = \rho_{\uparrow}(\mathbf{r}) + \rho_{\downarrow}(\mathbf{r})$ being the electronic density. While it is usually stated that $E_{\text{xc}}$ is unknown, this is not fully correct: $E_{\text{xc}}$ can be split in its exchange ($E_{\text{x}}$) and correlation ($E_{\text{c}}$) contributions, being only the latter the one that is really unknown. From the exact Fock expression for $E_{\text{x}}$ one may obtain the corresponding exact-exchange (EE) \textit{local} potential, to be used in the KS effective single-particle equations\cite{GKKG00,KK08}. Along the years, many advantages of the EE formalism have been addressed. To quote just a few: correct asymptotic behavior of $v_{\text{KS}}^{\sigma}(\mathbf{r})$ for atoms and molecules\cite{GKKG00}, and for solid surfaces\cite{HPR06}, complete cancellation of the self-interaction error between
the Hartree energy and conventional density-based approximations for the exchange energy\cite{KK08}, and considerable improvement of the KS gaps of semiconductors\cite{SMMVG99}. 
The EE method also provides a natural solution to the hard problem that represent electronic systems of reduced dimensionality. Differently from the usual functionals based on local density approximations, in the EE scheme $E_{\text{x}}$ is an explicit functional of the KS orbitals, thus dimensionality of the system is automatically and rigorously included. A good example of this is the great improvement achieved in the computation of many-body effects on the electronic properties of quasi two-dimensional electron gases (q2DEG) using either the EE formalism\cite{RP03,RPR05} or the more elaborate Optimized Effective Potential (OEP) method, where correlation effects are also computed through an orbital-dependent $E_{\text c}$ energy functional\cite{RP06,RP07}.

Possible generalizations of the DFT formalism to the high magnetic field regime of the Fractional Quantum Hall effect (FQHE) have been proposed by Ferconi, Geller and Vignale\cite{FGV95} and by Heinonen, Lubin, and Johnson\cite{HLJ95}. The inherent fractional filling factors $\nu$ needed for the FQHE were generated in the first case by appealing to a finite-temperature DFT formalism, and through an ensemble DFT scheme in the second case. In both works, a suitable \textit{ad-hoc} $E_{\text{xc}}(\nu)$ with slope discontinuities at the main fractional $\nu$'s was generated. Very recently, a new DFT formalism for the FQHE has been presented, based in the composite-fermion description of the FQHE\cite{ZTJSJ17}.

Detailed Hartree-Fock calculations \cite{Giuliani1985, jungwirth2000pseudospin, Jungwirth2001,miravet2016} have established a classification of the ferromagnetic regimes within the Integer Quantum Hall effect (IQHE). Notably, the application of DFT methods to this regime has been much less considered and limited to local density approximations \cite{Freire2007, FFE10}, or solutions in limiting regimes\cite{Morbec2008PRB,Morbec2009}.
A possible reason for this is that the experiments within the IQHE \cite{Muraki2001, Ellenberger2006,Zhang2005,Zhang2006,Zhang2007,Guo2008,Guo2010,Gusev2003,Lara2012,Lara2014,Larkin2013} are usually satisfactorily explained by effective non-interacting models that captures the correct physics of fully filled Landau levels in terms of effective g-factors \cite{Ellenberger2006}, plus phenomenological descriptions of the localization effects induced by disorder\cite{WvK11}. While this is in principle correct as far as correlation effect concerns, we shown here that exchange interactions modifies considerably the single-particle description of the IQHE. These are particularly relevant at the crossings of Landau levels, where ferromagnetic phase transitions occur. The EE formalism presented here is ideally suited for achieve this goal.
The present DFT-EE scheme is formulated in a way that can be applied to any electronic system with translational symmetry in a plane. In particular, to a q2DEG confined in a semiconductor quantum well with modulation- or delta-doped barriers, and with a magnetic field $B$ applied along the perpendicular to the plane, as shown schematically in Fig.~1. 

\begin{figure}
\begin{center}
\includegraphics[width=8.6cm]{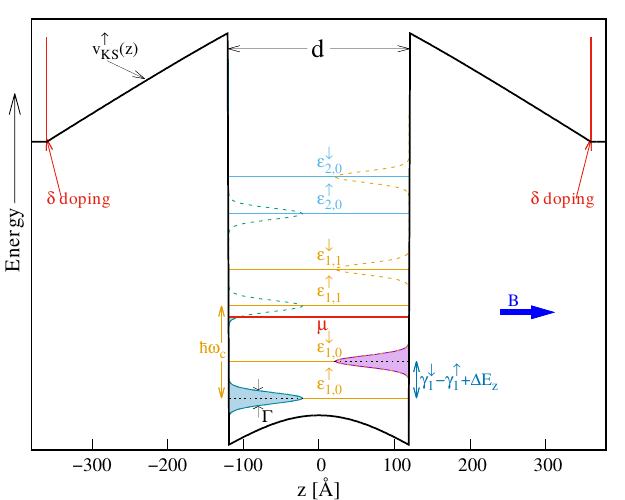}
\caption{Schematic view of the q2DEG for a GaAs quantum well of width $d = 240$\AA, and symmetric delta-doping layers on the $Al_xGa_{1-x}As$ barriers. $v_{\text{KS}}^{\sigma}(z)$ is the self-consistent KS potential, $\varepsilon_{i,n}^{\sigma}$ are the broadened Landau levels, and $\Delta E_z$ is the Zeeman splitting. Only the ground-state subband $i=1$ is occupied.}
\label{Fig1}
\end{center}
\end{figure}

\section{Kohn-Sham formulation}
Within the KS formulation of DFT, the 3D spin($\sigma$)-dependent KS orbitals of the present system in the Landau gauge may be factorized as $\Psi_{i,n,k}^{\sigma}(x,y,z) = \phi_{n}(x)e^{iky}\lambda_i^{\sigma}(z)/\sqrt{L}$, where 

\begin{equation}
 \phi_{n}(x) = \frac{\exp\left[-\frac{\left(x-l_B^{2}k\right)^{2}}{2l_B^{2}}\right]}
                 {\left[\sqrt{\pi} \, l_B \, 2^{n}\left(n!\right)\right]^{1/2}}
                 H_{n}\left(\frac{x-l_B^{2}k}{l_B}\right) \; ,
\end{equation}
and the $\lambda_i^{\sigma}(z)$ are the self-consistent solutions of the effective one-dimensional KS equations
\begin{equation}
\left[-\frac{1}{2}\frac{\partial^2}{\partial z^2}+v_{\text{KS}}^\sigma(z)\right]\lambda_i^{\sigma}(z)=\gamma_i^\sigma(\nu) \; \lambda_i^{\sigma}(z) \; ,
\end{equation}
in effective atomic units (effective Bohr radius $a_0^* = \epsilon \hbar^2/e^2 m^*$, and effective Hartree $Ha^* = m^* e^4 / \epsilon^2 \hbar^2$).
The full 3D eigenvalues associated with $\Psi_{i,n,k}^{\sigma}(x,y,z)$ are given by $\varepsilon_{i,n}^\sigma(\nu) =\gamma_i^{\sigma}(\nu) + (n+1/2)\hbar \omega_c/Ha^* -|g|\mu_B B s(\sigma) /(2 Ha^*$). Here $\omega_c = eB/m^*c$ is the cyclotron frequency, and the last term is the Zeeman coupling, with $s(\uparrow)=+1$, and $s(\downarrow)=-1$. The set of energy levels $\varepsilon_{i,n}^\sigma(\nu)$ are the Landau levels (LL) of the q2DEG. Each LL is represented by a Gaussian density of states (DOS) of half-width $\Gamma$ (see Fig.~1), that represents the disorder effects from charged impurities, interface defects, etc.\cite{AU74,B88}. 
$H_n(x)$ are the $n$-th Hermite polynomials, and $n \; (= 0,1,2,...)$ is the orbital quantum number
index. $k$ is the one-dimensional wave-vector label that distinguishes states within a given LL, each with a degeneracy $N_\phi = AB/\Phi_0$. $A$ is the area of the q2DEG in the $x-y$ plane, $B$ is the magnetic field strength in the $z$ direction, and $\Phi_0= ch/e$ is the magnetic flux number.   $l_B = \sqrt{c\hbar/eB}/a_0^*$ is the magnetic length.  $\nu=N/N_{\phi}$ is the dimensionless filling factor, with $N$ being the total number of electrons. The $\lambda_i^{\sigma}(z)$ are the self-consistent KS eigenfunctions for electrons in subband $i \; (= 1,2,...)$, spin $\sigma \; (=\uparrow,\downarrow)$  and eigenvalue $\gamma_i^\sigma(\nu)$. The spin-dependent KS potential is the sum of three contributions: 
$v_{\text{KS}}^\sigma = v_{\text{ext}}(z)+v_{\text{H}}(z)+v_{\text{xc}}^{\sigma}(z)$. $v_{\text{ext}}(z)$ represents the epitaxial potential plus the external fields (in the present case the electric field generated by the delta-doping layers).
$v_{\text{H}}(z)$ is the classical Hartree potential. $v_{\text{xc}}^{\sigma}(z)$ is the \textit{local} exchange-correlation (xc) potential to be defined below. 

The xc potential can be further split in the form $v_{\text{xc}}^{\sigma}(z)=v_{\text{x}}^{\sigma}(z)+v_{\text{c}}^{\sigma}(z)$. In the following we shall consider only its exchange contribution, assuming that correlation effects are negligible for $\nu$'s around integer values (IQHE), since the phase-space blocking situation of a set of full Landau levels precludes the existence of low-energy correlation-induced effects. In other words, as the number of electrons on the LL's is close to its full occupation, only a few Slater determinants would contribute to the wave-function, thus the exchange energy dominates over the correlation effects. This amounts to the ``exact-exchange'' characterization of the present computational approach.

\section{Exact-exchange at finite magnetic field}
The exact 3D Fock expression for the exchange energy is given by 
\begin{align}
E_{\text{x}}= & -\sum_{a,b,\sigma} f_a^\sigma f_b^\sigma \int d^{3}r \int d^{3}r' 
\frac{\Psi_a^\sigma(\mathbf{r})^*\Psi_b^\sigma(\mathbf{r}')^*\Psi_b^\sigma(\mathbf{r})\Psi_a^\sigma(\mathbf{r}')}{2 \; |\mathbf{r}-\mathbf{r}'|} \; ,\label{eq:Exact_exchange_energy}
\end{align}
with $f_a^\sigma,f_b^\sigma$ being the finite temperature weights, taking values between 0 and 1.
Substituting the eigenfunctions $\Psi_{i,n,k}^{\sigma}(\mathbf{r})$ in Eq.~(\ref{eq:Exact_exchange_energy}) and using the quasi-2D Fourier representation of the Coulomb interaction\cite{B88}, we obtain 
\begin{align}
E_{\text{x}}= & -\frac{N_\phi}{2l_B}\sum_{i,j,\sigma}\int dz\int dz' \; \lambda_{i}^{\sigma}(z)\lambda_{i}^{\sigma}(z')\lambda_{j}^{\sigma}(z)\lambda_{j}^{\sigma}(z')\nonumber \\
 & \times 
\sum_{n,m}n_{i,n,\sigma}^{\text{2D}}n_{j,m,\sigma}^{\text{2D}}I_{n}^{m}(z-z') \; .
\label{eq:Exact_exchange_energy_B_field}
\end{align}
Here $n_{i,n,\sigma}^{\text{2D}} = \int g(\epsilon-\epsilon_{i,n}^{\sigma})f_{\text{FD}}(\epsilon)d\epsilon$ is the occupation factor of a LL labeled by {$i,n,\sigma$}. $f_{\text{FD}}(\epsilon) = [1+e^{(\epsilon-\mu)/(k_BT)}]^{-1}$  is the Fermi-Dirac distribution function, and $\mu$ is the chemical potential. $g(\epsilon)$ is the DOS normalized to $1$, so that $0\leq n_{i,n,\sigma}^{\text{2D}} \leq 1$. Under these conditions, $\sum_{i,n,\sigma} n_{i,n,\sigma}^{\text{2D}} = \nu$ is constant and defines $\mu$. Finally,
\begin{align}
I_{n}^{m}(z-z') & = \frac{n_{<}!}{n_{>}!}\int_{0}^{\infty} dx \; e^{-x^{2}/2} ({x^{2}}/{2})^{n_{>}-n_{<}} \nonumber \\
 & \times \left[L_{n<}^{n_{>}-n_{<}}({x^{2}}/{2})\right]^{2}e^{-{x|z-z'|}/{l_{B}}}, 
\end{align}
as given elsewhere\cite{miravet2016}.
Here $L_n^m(x)$ are the generalized Laguerre polynomials, and $n^< = \min(n,m)$, $n^> = \max(n,m)$.
We emphasize that Eq.(\ref{eq:Exact_exchange_energy_B_field}) is valid for \textit{all} values of $\nu$, either integer or fractional, 
which results from an ``ensemble average'' defined as follows. While in a perfect crystal the degenerate states within a LL can be labeled by $k$ and a delta-Dirac DOS, an average over an ensemble of impurities and defects breaks this degeneracy, yielding the Gaussian DOS used above. This allow us to convert sums over the occupied $k$ states into the integral defined in $n_{i,n,\sigma}^\text{2D}$.

%\red{This results from an ``ensemble average'', which replaces the sum over the degenerate $k$ values of a pristine LL with the %integral over the Gaussian DOS above.}

When only the first subband $i=1$ is occupied, Eq.(\ref{eq:Exact_exchange_energy_B_field}) simplifies drastically to

%\begin{equation}
% e_{\text{x}}(\nu) = \frac{-1}{2 \nu l_B }\sum_{\sigma}\int dz\int dz'\rho_{\sigma}(z)\rho_{\sigma}(z') 
% \; \frac{S_{1}^{\nu_{\sigma}}(\Delta z)}{{(\nu_{\sigma} N_\phi)}^{2}} \; ,
%\label{eq:Ex_one subband}
%\end{equation}
\begin{equation}
 e_{\text{x}}(\nu) = \frac{-1}{2 \nu l_B }\sum_{\sigma} 
 \frac{\left\langle \rho_{\sigma} | S_{1}^{\nu_{\sigma}} |\rho_{\sigma} \right\rangle}{(\nu_{\sigma} N_\phi)^{2}} \; , 
% \; \frac{S_{1}^{\nu_{\sigma}}(\Delta z)}{{} \; ,
 \label{eq:Ex_one subband}
\end{equation}
with $e_{\text x} = E_{\text x}/N$ being the exchange energy per particle \cite{ex3d-comment},
$\left\langle \rho_{\sigma} | S_{1}^{\nu_{\sigma}} |\rho_{\sigma} \right\rangle$ representing integrals over $z$ and $z'$ of the densities $\rho_{\sigma}(z)$ and $\rho_{\sigma}(z')$ times $S_{1}^{\nu_{\sigma}}(z-z')$, 
and

\begin{equation}
 S_{1}^{\nu_{\sigma}}(z-z') = \sum_{n,m} n_{1,n,\sigma}^{\text{2D}} \; n_{1,m,\sigma}^{\text{2D}} \; I_{n}^{m}(z-z') \; .
 \label{eq:sum}
\end{equation}
Here we have used that $\rho_{\sigma}(z)  = N_{\sigma} |\lambda_1^{\sigma}(z)|^{2} = \nu_{\sigma} N_\phi |\lambda_1^{\sigma}(z)|^{2}$, and that $\nu_\sigma=\sum_{n}n_{1,n,\sigma}^{\text{2D}}$ is the spin-dependent filling  factor. 
Since $\int \rho_{\sigma}(z) dz = \nu_{\sigma} N_{\phi}$, this allow us to define $\nu N_{\phi} /A^* = (\pi r_s^2)^{-1}$, with $r_s$ being the 2D dimensionless parameter that characterizes the electronic in-plane density, and $A^*$ is the area in units of $a_0^*$. Using these relations, one obtains $\nu = 2 (l_B/r_s)^2$. This one-subband ($1S$) regime is quite relevant both from the experimental and theoretical viewpoints. Real q2DEG are easily driven to this regime by a suitable modulated or delta-doping design, bias application, or by changing the width $d$ of the quantum well, barrier height, etc.\cite{B88}. From the theoretical side, it was realized already some time ago the considerable simplification one gets in the EE defining equations at zero-magnetic field, when restricted to this regime \cite{RP03,RPR05,RP06,RP07,RHP15,N16}. In particular, in the $B=0$ $1S$ regime the EE potential is given by an explicit analytical expression, while in the many-subband regime $i > 1$ it is defined through an integral equation that must be solved numerically. As shown below, all these nice features of the $1S$ regime are preserved in the IQHE situation, even for large values of the filling factor $\nu$. 

The expression for $e_{\text x}(\nu)$ may be further simplified if we suppose that the LL broadening $\Gamma$ is smaller than the energy difference between consecutive LL's with the same spin ($\hbar\omega_{c}>\Gamma$). Then, denoting by $[\nu_{\sigma}]$ the integer part of $\nu_{\sigma}$, the occupation factors are just given by 
\begin{align}
n_{1,n,\sigma}^{\text{2D}} \equiv n_{n,\sigma}^{\text{2D}} = & \begin{cases}
1 & n < \left[\nu_{\sigma}\right] \; \\
p_\sigma & n = \left[\nu_{\sigma}\right] \; \\
0 & n > \left[\nu_{\sigma}\right] \; ,
\end{cases}
 \label{of}
\end{align}
where  $p_\sigma=\nu_\sigma-\left[\nu_{\sigma}\right]$, and $0 < p_{\sigma} <1$ is the fractional occupation factor of the more energetic occupied LL with spin $\sigma$. This allow us to simplify the sum in Eq.(\ref{eq:sum}) as follows, 
\begin{equation}
 S_{1}^{\nu_{\sigma}}(t) = a^{\nu_{\sigma}}(t) + 2 \; p_{\sigma} \; b^{\nu_{\sigma}}(t) + p_{\sigma}^2 \; c^{\nu_{\sigma}}(t) \; , 
 \label{S_1}
\end{equation}
with $a^{\nu_{\sigma}}(t) = \sum_{n,m=0}^{[\nu_{\sigma}]-1}I_{n}^{m}(t)$, 
$b^{\nu_{\sigma}}(t) = \sum_{n=0}^{[\nu_{\sigma}]-1}I_{n}^{[\nu_{\sigma}]}(t)$, 
$c^{\nu_{\sigma}}(t) = I_{[\nu_{\sigma}]}^{[\nu_{\sigma}]}(t)$, and $t=z-z'$.
%\begin{align}
% S_{1}^{\nu_{\sigma}}(t)= &\sum_{n,m=0}^{[\nu_{\sigma}]-1}I_{n}^{m}(t) +
%                          2f_\sigma\sum_{n=0}^{[\nu_{\sigma}]-1}I_{n}^{[\nu_{\sigma}]}(t) +
%                          f_\sigma^{2} \; I_{[\nu_{\sigma}]}^{[\nu_{\sigma}]}(t).
% \label{S_1}
%\end{align}

When constrained to the strict 2D limit, which amounts to the replacement $|\lambda_1^{\sigma}(z)|^2 \rightarrow \delta(z)$,
Eq.~(\ref{eq:Ex_one subband}) with the $S_{1}^{\nu_{\sigma}}(t)$ as approximated in the last equation, exactly reduces to the result for the strict 2D exchange energy for fractional filling factors, as obtained by taking the low-temperature limit from the finite-temperature expression for $e_{\text{x}}(\nu)$.\footnote{See for example Eq.(10.93) in Ref.\cite{GV}.}
Each term in Eq.(\ref{S_1}), and also then in $e_{\text{x}}(\nu)$ admits a transparent physical interpretation. The first term represents the exchange energy associated with spin-$\sigma$ electrons in $[\nu_{\sigma}]-1$ fully filled LL's. The second, linear in $p_{\sigma}$, corresponds to the exchange energy due to the interaction between the electrons in the $[\nu_{\sigma}]$ LL and those in the lower levels. The third term, proportional to $p_{\sigma}^2$, represents the exchange energy among electrons in the partially occupied $[\nu_{\sigma}]$ LL. We note here that all numerical results to be presented below were obtained by using the full expression for $S_{1}^{\nu_{\sigma}}(t)$ in Eq.(\ref{eq:sum}); nonetheless the approximated Eq.(\ref{S_1}) is quite useful to understand the full numerical results. 

Next, we obtain the spin-dependent EE potential from 
$v_{\text x}^{\sigma}(z)= N {\delta e_{\text x}}/{\delta \rho_{\sigma}(z)}$ for the $1S$ case, which reads
\begin{align}
v_{\text x}^{\sigma}(z) = \frac{-1}{l_{B}N_{\phi} \nu_\sigma^2} \int dz'\rho_{\sigma}(z')S_{1}^{\nu_{\sigma}}(z-z') \nonumber \\
-\frac{1}{2 l_{B}N_{\phi}} \int dz\int dz'\rho_{\sigma}(z)\rho_{\sigma}(z')\nonumber \\
\times\frac{\partial\left(S_{1}^{\nu_{\sigma}}(z-z')/\nu_\sigma^2\right)}{\partial\gamma_{1}^{\sigma}}\frac{\partial\gamma_{1}^{\sigma}}{\partial\rho_{\sigma}(z)}.
 \label{inter}
\end{align}
where the first (second) term comes from the explicit (implicit) dependence of $E_{\text x}$ on $\rho_{\sigma}(z)$. This implicit dependence is easy of understand: by changing $\rho_{\sigma}(z)$, a change in $v_{\text H}(z)$ and $v_{\text x}^{\sigma}(z)$ is induced through the self-consistent solution of the KS equation, that in turn affects $\gamma_1^{\sigma}$. After some cumbersome calculations, Eq.(\ref{inter}) becomes
\begin{equation}
 v_{\text x}^{\sigma}(z) = u_{\text x}^{\sigma}(z) + \overline{\Delta v}_{\text x}^{\; \sigma} \; ,
 \label{eq:V_x}
\end{equation} 
with $u_{\text x}^{\sigma}(z)$ being the first term in Eq.(\ref{inter}),  
$\overline{\Delta v}_{\text x}^{\; \sigma} = \eta_{\text x}^{\nu_{\sigma}} - \bar{u}_{\text x}^{\sigma}$,
$\bar{u}_{\text x}^{\sigma} = \int \lambda_1^{\sigma}(z)^2 {u}_{\text x}^{\sigma}(z) dz$, and 
$\eta_{\text x}^{\nu_{\sigma}} = -\left\langle \rho_{\sigma} | S_2^{\nu_{\sigma}} | \rho_{\sigma} \right\rangle / 
\left(\nu_{\sigma}^2 (N_{\phi})^2 l_B \right)$. It remains to define
\begin{equation}
 S_{2}^{\nu_{\sigma}}(t) = \frac{\sum_{n,m}\left({\partial n_{n,\sigma}^{\text {2D}}}/{\partial\gamma_{1}^{\sigma}}\right)
 n_{m,\sigma}^{2D} \; I_{n}^{m}(t)}
 {\sum_{n}\left({\partial n_{n,\sigma}^{\text{2D}}}/{\partial\gamma_{1}^{\sigma}}\right)} \; .
 \label{eq:S2}
\end{equation}

Eqs.(\ref{eq:Ex_one subband}) and (\ref{eq:V_x}) are the main results of this paper. Notice that Eq.(\ref{eq:V_x}) is \textit{not} invariant under a constant shift $v_\text{x}^{\sigma}(z) \rightarrow v_\text{x}^{\sigma}(z) + C$, since here we consider the grand-canonical ensemble. That is, for given values of the external parameters $\nu$, $r_s$, and $T$, the chemical potential $\mu$ is fixed. A rigid shift in $v_\text{x}^{\sigma}(z)$ will induce then a rigid shift in the KS eigenvalues $\gamma_i^{\sigma}$, that at constant $\mu$ will modify the occupation factors $n_{n,\sigma}^{\text {2D}}$, leading to a change in $v_\text{x}^{\sigma}(z)$. In other words, $v_\text{x}^{\sigma}(z)$ is \textit{fully} determined by Eq.(\ref{eq:V_x}), and then no floating constant should be fixed by imposing asymptotic boundary conditions, as it is usual for similar closed systems\cite{RHP15,KPK09}.
Naturally, the KS equations must be solved numerically and self-consistently: For each $\nu$, both $v_{\text H}(z)$ and $v_{\text x}^{\sigma}(z)$ determine and are determined by the solutions $\lambda_1^{\sigma}(z)$ and $\gamma_1^{\sigma}$, which yields the self-consistent loop. The following set of $GaAs$ material parameters have been used in the numerical calculations: $m^*=0.067 m_0$, ($m_0$ being the bare electronic mass), $\epsilon = 12.85$, $g = -0.44$, $T = 340$ mK, and $\Gamma(B) = 0.150 \sqrt{B}$ meV. The $GaAs$ - $Al_xGa_{1-x}As$ conduction band barrier height has been taken as 228 meV, which corresponds to $x \simeq 0.3$.

\section{Results and discussion}
In the following we analyze the dependence upon $\nu$ of $e_{\text{x}}(\nu)$, and $v_{\text{x}}^{\sigma}(z)$. First, Fig.~2 shows the results for the exchange energy $e_{\text{x}}(\nu)$, as given by Eq.(\ref{eq:Ex_one subband}). 
$e_{\text x}(\nu)$ approach a constant asymptotic value in the limit of large $\nu$ (small $B$ limit). This is easy to understand: as $B \rightarrow 0$, at fixed density, electrons redistribute among an increasing number of LL's, such that the overlap of many Gaussian DOS reaches asymptotically the constant DOS of a q2DEG in the zero-field limit.

\begin{figure}
\includegraphics[width=8.6cm]{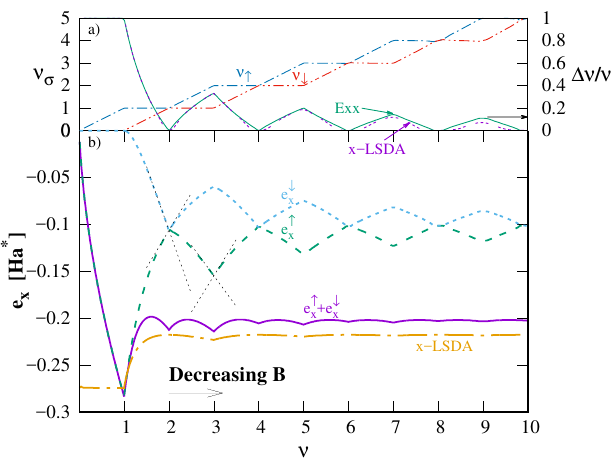}
\caption{
    (a) Spin-dependent filling factors $\nu_{\sigma}$ (left scale) and relative spin polarization $\Delta \nu / \nu$ (right scale) for increasing values of the total filling factor $\nu$; $\Delta \nu = \nu_{\uparrow} - \nu_{\downarrow}$. (b) Exchange energy per particle $e_{\text{x}}(\nu)$ versus $\nu$, at fixed density $r_s=2.5$. $e_{\text{x}}^{\uparrow}(\nu)$ and $e_{\text{x}}^{\downarrow}(\nu)$ are also shown. The straight lines at $\nu=2$ and $\nu=3$ represents the analytical approximations associated with the slope discontinuities of $e_{\text{x}}^{\downarrow}(\nu)$ and $e_{\text{x}}^{\uparrow}(\nu)$, respectively. The x-LSDA results are also shown for comparison.}
\label{Fig2}
\end{figure}

The oscillations of $e_{\text{x}}(\nu)$ are however its more interesting feature: At every integer value of $\nu$, $e_{\text{x}}(\nu)$ minimizes locally in a non-analytical way, yielding an inverted ``cusp''. Besides, at each odd $\nu \; (=\nu^o)$,
$e_{\text{x}}^{\uparrow}(\nu)$ has a local minimum while $e_{\text{x}}^{\downarrow}(\nu)$ exhibits a local maximum, with the sum 
of both resulting in a weaker local minimum in $e_{\text{x}}(\nu)$, due to a partial cancellation between the two opposite
behaviours. An equivalent situation happens at each even $\nu \; (=\nu^e)$, but with the roles of $e_{\text{x}}^{\uparrow}(\nu)$ 
and $e_{\text{x}}^{\downarrow}(\nu)$ exchanged. The qualitative behaviour of $e_{\text{x}}^{\uparrow}(\nu)$ and $e_{\text{x}}^{\downarrow}(\nu)$ around each $\nu^o$ and $\nu^e$ is easy of understand. For example, at each $\nu^o$ the relative spin-polarization displayed in the upper panel of Fig.~2 attains its possible maximum value $\Delta \nu / \nu = 1 / \nu^o$, and this optimizes the spin-up exchange energy gain. However, for $e_{\text{x}}^{\downarrow}(\nu)$ this is the worst possible configuration, and then it exhibits a local maximum. For each $\nu^e$, on the other side, $\Delta \nu / \nu = 0$ as corresponds to a spin-compensated situation, and this optimizes the gain of the spin-down exchange energy, but delivers the smallest energy gain in $e_{\text{x}}^{\uparrow}(\nu)$, that displays thus a local maximum. Besides these qualitative considerations, one can obtain the same results analytically, starting directly from Eq.(\ref{eq:Ex_one subband}) and expanding around every $\nu^o$ and $\nu^e$. Proceeding this way, we have found that both $e_{\text{x}}^{\uparrow}(\nu)$ and $e_{\text{x}}^{\downarrow}(\nu)$ depart linearly from each integer $\nu$, and these analytical linear approximations are represented by the crossing straight lines displayed at 
$\nu^e = 2$, and $\nu^o = 3$.

The x-LSDA results are also shown in Fig.~2. The exchange energy per particle is obtained from the corresponding expression for the spin-polarized homogeneous interacting 3D electron gas\cite{GV}. 
%It depends on the density and the degree of spin-polarization, and displays weak oscillations around odd filling factors, where the spin-polarization reaches a local maximum value. 
For large $\nu$ (small $B$), the x-LSDA overestimates the EE energy, but this is not a general trend. For bigger densities (smaller $r_s$), we have found that it underestimates the EE energy (not shown).Interestingly, the x-LSDA displays small oscillations with non-analytical cusps around odd $\nu$, where the spin polarization [Fig.~\ref{Fig2}(a)] is maximum.

\begin{figure}
\includegraphics[width=8.6cm]{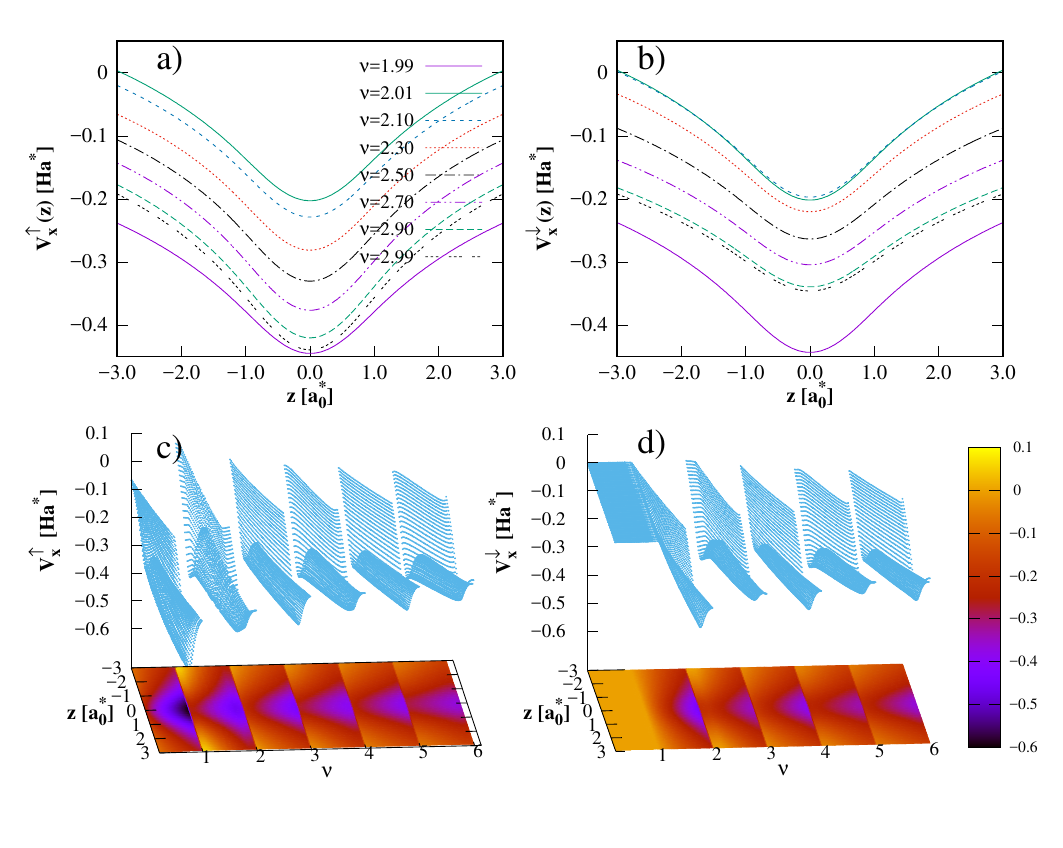}
\caption{Exact-exchange potentials $v_\text{x}^{\uparrow}(z)$ and $v_\text{x}^{\downarrow}(z)$ as a function of $z$ for selected values of $\nu$ (upper panels) around $\nu = 2$, and as a function of $z$ and $\nu$ (lower panels). Graded colors in the $z-\nu$ plane corresponds to a bidimensional projection of the EE potentials. Both EE potentials suffer abrupt discontinuous jumps at every integer $\nu$. The density is fixed at $r_s = 2.5$.}
\label{Fig3}
\end{figure}

The remarkable behavior of $v_{\text x}^{\uparrow}(z)$ and $v_{\text x}^{\downarrow}(z)$ as a function of $z$ and for several values of $\nu$ is displayed in Fig.~\ref{Fig3}.  
The crucial feature to note from Fig.~\ref{Fig3} is the abrupt jump (a rigid upward shift) of $v_\text{x}^{\uparrow}(z)$ and $v_\text{x}^{\downarrow}(z)$ when $\nu$ crosses an integer value.
After the jump, and as the filling of the corresponding LL proceeds, both $v_\text{x}^{\uparrow}(z)$ and $v_\text{x}^{\downarrow}(z)$ somehow follow the opposite behavior, and start to move down in a continuously. As shown in the upper panel, for $\nu=2.99$ both EE potentials are again close to their value at $\nu=1.99$, just before the jump. It should be noted here, however, that only for $p_{\sigma} \rightarrow 0$, the difference $v_\text{x}^{\sigma}(z)[\nu=[\nu]+p_{\sigma}] - v_\text{x}^{\sigma}(z)[\nu=[\nu]-p_{\sigma}]$ becomes a constant. For finite (not infinitesimal) $p_{\sigma}$, the potentials at filling factors smaller and greater than $[ \nu ]$ differ by a $z$-dependent function, as can be seen in Fig.~\ref{Fig3}.

\begin{figure}
 \includegraphics[width=8.6cm]{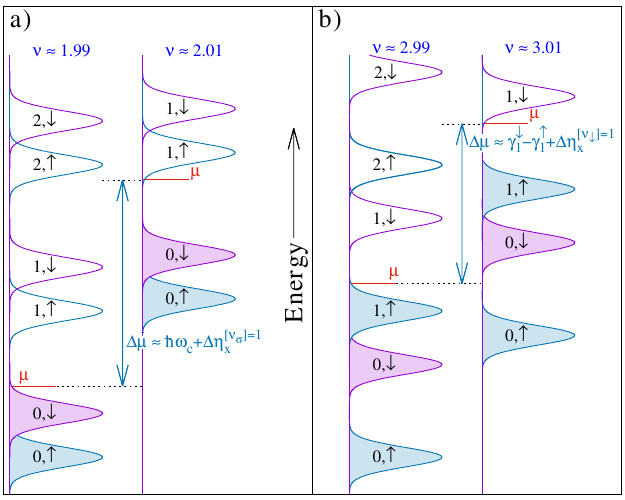}
 \caption{Schematic view of the lowest energy LL: a) $\nu$ close to $\nu=2$, and b) $\nu$ close $\nu=3$. The straight horizontal line corresponds to different positions of the chemical potential $\mu$.}
 \label{Fig4}
\end{figure}

The discontinuity in $v_{\text{x}}^{\sigma}(z)$ at integer values of $\nu$ can be traced back to a discontinuity in 
$S_{2}^{[\nu_{\sigma}]}(t)$, that in turn induces a discontinuity in $\eta_\text{x}^{[\nu_{\sigma}]}$. To explain this, we should return to the general expression for $S_{2}^{[\nu_{\sigma}]}(t)$ in Eq.(\ref{eq:S2}), and assume that for $\nu$ close to an integer value (either even or odd) essentially only one (per spin) of the occupation factors $n_{n,\sigma}^{\text{2D}}$ contributes to the derivative with respect to $\gamma_1^{\sigma}$. This situation is schematically depicted in Fig.~\ref{Fig4}, for fillings close to $\nu=2$ (left panel) and close to $\nu=3$ (right panel). Expressed in a different way, the assumption amounts to the approximation 
${\partial n_{n,\sigma}^{\text{2D}}} / {\partial\gamma_{1}^{\sigma}} \simeq \delta_{n,[\nu_{\sigma}]} \; ({\partial n_{[\nu_{\sigma}],\sigma}^{\text{2D}}} / {\partial\gamma_{1}^{\sigma}} ) $. Referring to the situation displayed in Fig.~\ref{Fig4}, for $\nu \simeq 1.99$, $[ \nu_{\uparrow} ] = [ \nu_{\downarrow} ] = 0$, and these are the only two LL that contributes to $S_{2}^{[\nu_{\sigma}]}(t)$ in Eq.(\ref{eq:S2}) (one for each spin); for $\nu \simeq 2.01$ instead, $[ \nu_{\uparrow} ] = [ \nu_{\downarrow} ] = 1$, and only for them the occupation factors change when $\gamma_1^{\sigma}$ changes. 

\begin{figure}
 \includegraphics[width=8.6cm]{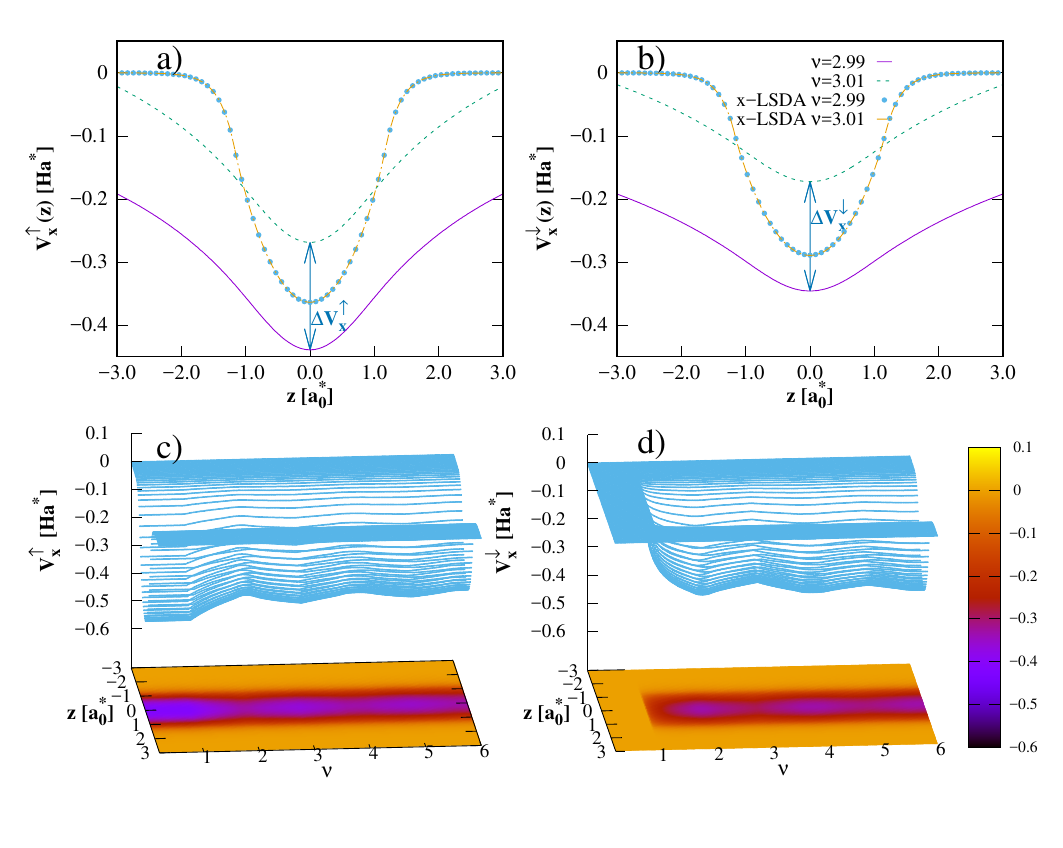}
 \caption{x-LSDA potentials $v_\text{x}^{\uparrow}(z)$ and $v_\text{x}^{\downarrow}(z)$ as a function of $z$ for selected values of $\nu$ (upper panel) around $\nu = 3$, and as a function of $z$ and $\nu$ (lower panel). Graded colors in the $z-\nu$ plane corresponds to a bidimensional projection of the corresponding potentials. The x-LSDA results are continuous as $\nu$ passes through integer values. The density is fixed at
$r_s = 2.5$.}
 \label{Fig5}
\end{figure}

Proceeding in this way, Eq.(\ref{eq:S2}) may be simplified to 
\begin{equation}
 S_{2}^{[\nu_{\sigma}]}(t) \simeq \sum_n n_{n,\sigma}^{\text{2D}} \; I_n^{[\nu_{\sigma}]}(t) \; .
 \label{eq:S2*}
\end{equation}
We emphasize here that Eq.(\ref{eq:S2*}) is valid only for $\nu$ close to integer values, while Eq.(\ref{eq:S2}) is valid for arbitrary filling factors. Application of Eq.(\ref{eq:S2*}) to the filling factors displayed in Fig. \ref{Fig4} leads to the same result in all cases. For the discontinuity in $S_{2}^{[\nu_{\sigma}]}$ we obtain thus
\begin{eqnarray}
 \Delta S_{2}^{[\nu_{\sigma}]}(t) & \equiv & S_{2}^{[\nu_{\sigma}]+p_{\sigma}}(t) - S_{2}^{[\nu_{\sigma}]-p_{\sigma}}(t) \nonumber \\
                            &=& \sum_{n=0}^{[\nu_{\sigma}]-1} 
 \left( I_n^{[\nu_{\sigma}]}(t) - I_n^{[\nu_{\sigma}]-1}(t)\right) < 0\; ,
 \label{disc}  
\end{eqnarray}
and the negative sign is justified below for a few simple cases.
From this, one obtains the discontinuity in $\eta_{\text x}^{[\nu_{\sigma}]}$,
\begin{eqnarray}
 \Delta \eta_{\text x}^{[\nu_{\sigma}]} \equiv  
 \eta_\text{x}^{[\nu_{\sigma}]+p_{\sigma}} - \eta_\text{x}^{[\nu_{\sigma}]-p_{\sigma}} \nonumber 
 = - \frac{\langle \rho_{\sigma} | \Delta S_{2}^{[\nu_{\sigma}]} | \rho_{\sigma} \rangle}{\nu_{\sigma}^2 (N_{\phi})^2 l_B}  >  0\; , \nonumber \\
 = \frac{-1}{r_s \sqrt{\nu/2}} \iint dz \, dz' (\lambda_1^{\sigma}(z) \lambda_1^{\sigma}(z'))^2 \Delta S_2^{[\nu_{\sigma}]}(z-z'),
 \label{eq:diseta}
\end{eqnarray}
with a positive sign, in agreement with the full numerical results displayed in Fig.~3.

The simplest case to analyze is the discontinuity at $\nu = 1$. In this case, $[\nu_{\uparrow}]=1$, $[\nu_{\downarrow}]=0$ and Eq.(\ref{disc}) reduces to
\begin{equation}
 \Delta S_{2}^{[\nu_{\uparrow}]=1}(t) = I_0^1(t) - I_0^0(t) \; ,
 \label{eq:nu1}
\end{equation}     
which admits a simple physical interpretation. $I_0^0(t)$ is proportional to the exchange energy among electrons in the filled ground-state LL (intra-LL exchange interaction), while $I_0^1(t)$ is proportional to the exchange interaction among electrons in the ground and first-excited LL's (inter-LL exchange interaction). Since usually intra-LL interactions are stronger than inter-LL ones, $I_0^0(t) > I_0^1(t)$ for all values of $t$, one gets the negative sign of $\Delta S_{2}^{[\nu_{\uparrow}]=1}$, that in turn leads to the abrupt positive jump in $v_x^{\uparrow}(z)$ displayed in Fig. 3 at $\nu=1$.
The next simple case is the discontinuity at $\nu=2$. Here $[\nu_{\uparrow}]=[\nu_{\downarrow}]=1$, with the result that
$\Delta S_{2}^{[\nu_{\uparrow}]=1}(t)=\Delta S_{2}^{[\nu_{\downarrow}]=1}(t)=I_0^1(t) - I_0^0(t)$, the same as above.
This implies that the discontinuities in $v_{x}^{\uparrow}(z)$ and $v_{x}^{\downarrow}(z)$ are the \textit{same} at $\nu=2$,
as observed in Fig.~3 and as expected from physical grounds. Note, however, that since $\Delta \eta_{\text x}^{[\nu_{\sigma}]}$
scales as $\nu^{-1/2}$ at fixed $r_s$ (see Eq.(\ref{eq:diseta})), the discontinuity at $\nu=2$ in $v_{\text x}^{\uparrow}(z)$ and $v_{\text x}^{\downarrow}(z)$ is approximately a factor $1/\sqrt{2}$ smaller than the discontinuity in $v_{\text x}^{\uparrow}(z)$ at $\nu=1$.
The next interesting case is $\nu=3$. Here $[\nu_{\uparrow}]=2$ and $[\nu_{\downarrow}]=1$, and then
\begin{equation}
 \Delta S_{2}^{[\nu_{\uparrow}]=2}(t) = I_0^2(t) + I_1^2(t) - I_0^1(t) - I_1^1(t) \; ,
\end{equation}
while $\Delta S_{2}^{[\nu_{\downarrow}]=1}(t)$ is once again given by Eq.(\ref{eq:nu1}). When replaced in Eq.(\ref{eq:diseta}), this leads to discontinuities of different sizes for $v_{\text x}^{\uparrow}(z)$ and $v_{\text x}^{\downarrow}(z)$ at $\nu=3$, as can be barely appreciated from the upper panels in Fig.~5; we also provide there a comparison between the EE and the x-LSDA potentials. The difference between the x-LSDA exchange potentials are indistinguishable on the scale of the drawing for $\nu=2.99$ and $\nu=3.01$, as expected. It is also worth of note the drastically different asymptotic behavior of both types of exchange potentials: the well-known exponential decay of x-LSDA should be contrasted with the much slower and physically correct $-1/z$ decay of the EE potentials. In the lower panel of Fig.~5 we give a global view of the x-LSDA exchange potential as a function of $z$ and $\nu$. Being just proportional to $(\rho_{\sigma}(z))^{4/3}$, the same has no discontinuities at integer values of $\nu$, showing instead changes in the slope as $\nu$ crosses integer values. This is also easily grasped from the projection in the $\nu - z$ plane. The much faster decay of the x-LSDA exchange potential is evident from the strong narrowing of the central segment, whose darkness is proportional to the deepness of the potential.

In retrospective, it is important to realize that the discontinuity originates from the $\eta_{\text x}^{\nu_{\sigma}}$ term, which comes  from the implicit derivative of the exchange energy with respect to the density. This is the term that includes the in-plane density ($\nu_{\sigma} N_{\phi} / A^*$) dependence of the exchange energy. The inclusion of the implicit derivative  is also important to recover the correct strict-2D limit at zero magnetic field\footnote{To be published.}. 

The six lowest LL eigenvalues $\varepsilon_{1,n}^{\sigma}(\nu) (\equiv \varepsilon_{n}^{\sigma}(\nu))$ are shown in Fig.~6, both in the EE and x-LSDA approaches. The discontinuities that the EE $v_{\text x}^{\uparrow}(z)$ and $v_{\text x}^{\downarrow}(z)$ have at integer values of $\nu$ induces an abrupt change of all the LL energies, and of the chemical potential $\mu$. These numerical self-consistent results fully validate the strong renormalization of the LL electronic structure schematically displayed in Fig.~4. Focusing first in the situation at $\nu \simeq 2$, the "doublet" LL ordering $\{ \varepsilon_{n}^{\uparrow} \; \varepsilon_{n}^{\downarrow} \}$ is easily observed from Fig.~6, and is schematically shown in the left panel of Fig.~4. The LL ordering changes however drastically for $\nu \simeq 3$, since now the "doublet" structure is built with 
$\{ \varepsilon_{n}^{\downarrow} \; \varepsilon_{n+1}^{\uparrow} \}$ pairs of LL, as shown schematically in the right panel of Fig.~4. How does the q2DEG passes from one configuration close to even $\nu$ to the case close to odd $\nu$?
The answer is in the evolution of $v_{\text x}^{\uparrow}(z)$ and $v_{\text x}^{\downarrow}(z)$ for $2.01 \leq \nu \leq 2.99$ displayed in the upper panel of Fig.~3. It is seen there how the continuous downward shift of $v_{\text x}^{\uparrow}(z)$ is larger than the one for $v_{\text x}^{\downarrow}(z)$, being the net result a global downward shift of the spin-up LL relative to the spin-down LL. And this is precisely the situation schematically depicted in Fig.~4, restricted however to the two limiting values $\nu = 2.01, \; 2.99$. Once again, these qualitative considerations are validated by inspection of how the self-consistent LL eigenvalues       
$\varepsilon_{n}^{\sigma}(\nu)$ evolve in Fig.~6 in this filling factor window.

The x-LSDA eigenvalues displayed in the right panel of Fig.~6 behaves in a quite different way. First, since the x-LSDA exchange potentials are continuous functions of $\nu$, no discontinuities are present in the corresponding eigenvalues. Second, the chemical potential $\mu$ still have however some broadened discontinuities, either proportional to the "cyclotron gap" $\hbar \omega_c$ at $\nu =$ even, or proportional to the  ''spin-gap" 
$\varepsilon_{n}^{\downarrow}(\nu) - \varepsilon_{n}^{\downarrow}(\nu)$ at $\nu =$ odd. For instance, $\hbar \omega_c \simeq$ 0.16 and 0.08 for $\nu=$ 2 and 4, respectively, in reasonable agreement with the jump in $\mu$ at these filling factors. Third, note that the jump in $\mu$ at $\nu=4$ (cyclotron gap) is bigger than the jump at $\nu=3$ (spin-gap).

%\begin{figure*}
%\begin{center}
%\includegraphics[width=16cm]{Figure_6.pdf}

\begin{figure}
 \includegraphics[width=8.6cm]{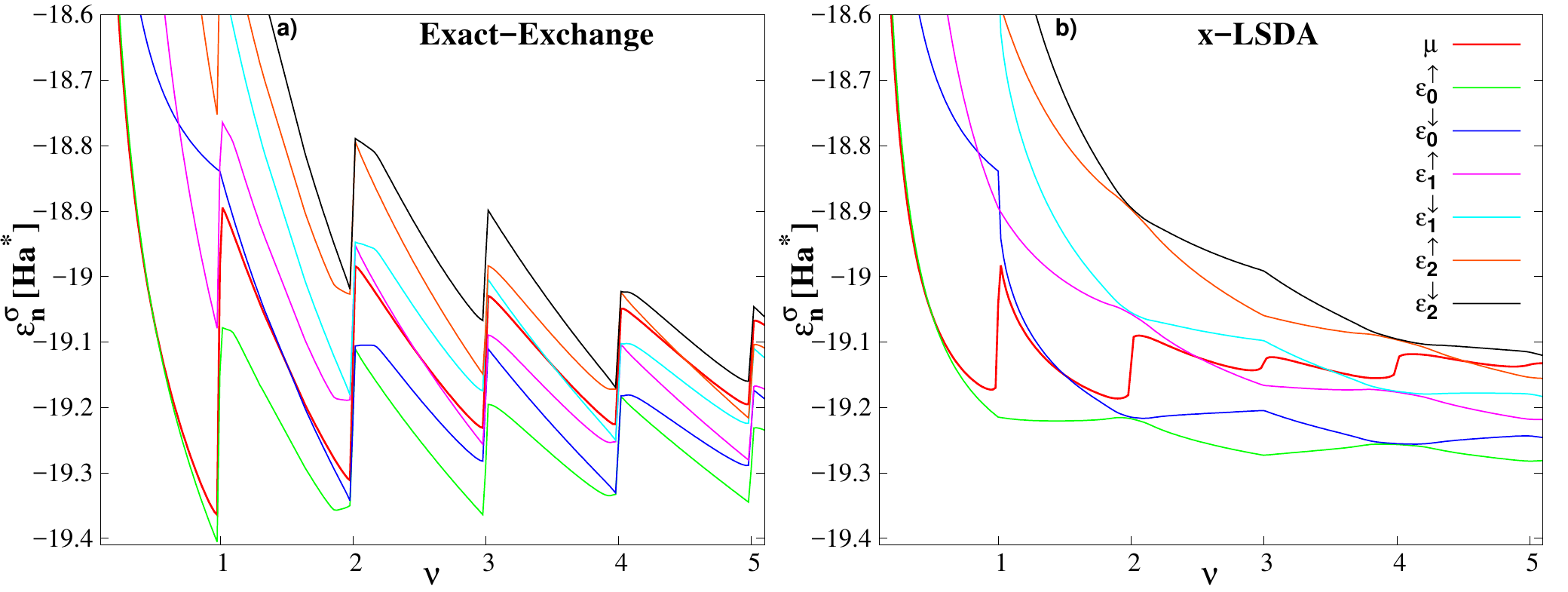}
\caption{Six lowest energy LL as a function of the filling factor $\nu$. a) Exact-exchange results, and b) x-LSDA results. The chemical potential $\mu$ is denoted with the thick red line. $r_s=2.5$ (fixed density).}
\label{Fig6}
\end{figure}
%\end{center}
%\end{figure*}

\section{Conclusions}
The exact-exchange energy functional of the integer quantum Hall effect (IQHE) has been found. It minimizes locally with discontinuities in the derivative at each integer filling factor $\nu$. In the one-subband regime, where only the ground-state subband of the quasi two-dimensional electron gas is occupied, an explicit analytical expression has been found for the associated spin-dependent exact-exchange potential. Its striking feature is that it jumps abruptly by a positive constant every time $\nu$ passes through an integer value. This is in analogy to the discontinuities in finite systems and solids when the total
number of electrons $N$ passes through integer values, but in our case the novelty is that the discontinuities are induced by the magnetic field, at fixed density. The size of the jump is the same for spin-compensated situations at each $\nu=$ even. For $\nu=$ odd, the discontinuities are different, being the jump bigger for the exact-exchange potential of the majority spin-component.
Strong differences are found regarding the standard x-LSDA for filling factors close to integer values.
%\section{2D limits}

%All results previously obtained  can be particularized to the 2D system. We obtain this  taken the orbital as %$\xi^\sigma(z)=\sqrt{\delta(z)}$, where $\delta(z)$ is the Dirac function. Using Laguerre polynomial relations we obtain
%\begin{align}
%\epsilon_x= &-\frac{1}{2\nu %l_B^\ast}\sum_\sigma\left(I_1([\nu_\sigma])+2I_2([\nu_\sigma])f_\sigma+I_3([\nu_\sigma])f_\sigma^2\right),
%\label{eq:ex-energy2D}
%\end{align} 

%\begin{align}
%\mu^{\sigma}_x(z)= & -\frac{1}{\nu_\sigma l_B^\ast} S_1^{\nu_\sigma}(|z|),
%\end{align} 

%\begin{align}
%\bar\mu^{\sigma}_x=\mu^{\sigma}_x(0)= & -\frac{1}{\nu_\sigma l_B^\ast} %\left(I_1([\nu_\sigma])+2I_2([\nu_\sigma])f_\sigma+I_3([\nu_\sigma])f_\sigma^2\right),
%\end{align}
%and 

%\begin{align}
%\eta^{\sigma}_x= & -\frac{1}{l_B^\ast} \left(I_2([\nu_\sigma])+I_3([\nu_\sigma])f_\sigma\right),
%\end{align}
%where 
%\begin{align}
%I_1(n)=\int_0^\infty e^{-x^2/2}[L_n^1(\frac{x^2}{2})]^2dx,
%\end{align}

%\begin{align}
%I_2(n)=\int_0^\infty e^{-x^2/2}L_n^0(\frac{x^2}{2})L_{n-1}^1(\frac{x^2}{2})dx,
%\end{align}

%\begin{align}
%I_3(n)=\int_0^\infty e^{-x^2/2}[L_n^0(\frac{x^2}{2})]^2dx.
%\end{align}

%the  expression (\ref{eq:ex-energy2D}) for exchange energy  reproduce  the result presented in \cite{Giuliani-Vignale} and in the %limits of $\nu\rightarrow\infty$ take the form of 2D-LDA expression for exchange energy. This 2D potential can be applied to %multilayer systems like  bilayer \cite{JM-2000} and trilayer \cite{MP-2016} in presence of magnetic field with the advantage of %the inclusion  of  intra-layer and inter-layer exchange naturally. 
\acknowledgments
D.M. and C.R.P. acknowledges the support of ANPCyT under Grant PICT-2012-0379. C.R.P.  thanks
%Consejo Nacional de Investigaciones Cient\'ificas y T\'ecnicas 
CONICET for partial financial support, grant PIP 2014-2016 00402.

\end{document}